\pdfoutput=1
\documentclass[runningheads]{llncs}

\usepackage[T1]{fontenc}
\usepackage{amsmath,amssymb}
\usepackage{booktabs}
\usepackage{tikz}
\usetikzlibrary{positioning,arrows.meta,calc}
\usepackage[hidelinks]{hyperref}

\newif\ifdraftnotes

\newcommand{\Prob}{\mathbb{P}}
\newcommand{\E}{\mathbb{E}}

\draftnotesfalse

\begin{document}

\title{Multi-Agent Screening: What Agreement and Acceptance Cannot Tell You}
\titlerunning{Multi-Agent Screening}

\author{
  Benjamin~Chang\inst{1}\orcidID{0009-0006-0332-0319} 
  \and
  Dell~Zhang\inst{2}\thanks{Corresponding author.}\orcidID{0000-0002-8774-3725}
}
\authorrunning{B. Chang and D. Zhang}
\institute{
  University of Cambridge, Cambridge, UK\\ 
  \email{btyc3@cam.ac.uk} 
  \and
  XSci, London, UK\\ 
  \email{dell.z@ieee.org}
}

\maketitle

\begin{abstract}
  When AI agents screen items for a costly human assessor, as in agentic search, e-discovery and AI reviewing of papers, only what they pass gets judged. 
We show what each kind of evidence can establish about their errors, even with unlimited data. 
Agreement shows that agents concur, not that they are right, however many there are. 
Acceptance by the assessor gives precision but only a lower bound on recall. 
Knowing which agents found which relevant items gives an upper bound on recall if agents' finds are not negatively correlated, and caps what one more agent can add if agents are interchangeable.
Without assumptions about the agents, recall can be pinned down only by judging items they did not pass: a random sample of these estimates how many relevant items were missed, to any chosen accuracy, and the sample needed does not grow with the number of agents.

\end{abstract}

\section{Introduction}
\label{sec:introduction}

Retrieval is becoming a multi-agent activity: agents rewrite queries, retrieve and rerank candidates, judge relevance~\cite{faggioli2023perspectives,thomas2024large} and review papers alongside or ahead of human experts~\cite{biswas2026aaai,bianchi2025agents4science}. 
These systems share a business problem: agents are cheap, but the assessor at the end, whether a relevance assessor, an expert or a lawyer, is costly, and so sees only what the agents pass. 
This is the \emph{incomplete-judgment} setting of pooled evaluation~\cite{buckley2004retrieval,zobel1998reliable} and e-discovery~\cite{webber2013approximate,lewis2021certifying}.

Operators face four questions. 
Can we trust the agents when they agree? 
Are we missing relevant items? 
Is one more agent worth paying for? 
How much assessment budget must we keep for checking the system? 
We ask what the system's own evidence can answer even with unlimited data, so the limits we find are not sampling error. 
Using classical tools adapted to \emph{multi-agent systems}~\cite{zhang2026masact}, we give a recall interval, a bound on what the next agent can add, and rules for spending on agents and audits, with AI reviewing of papers as a worked example.

\section{Multi-Agent Screening}
\label{sec:framework}

\begin{figure}[t]
  \centering
  \begin{tikzpicture}[
    font=\scriptsize, >={Stealth[length=1.5mm]}, node distance=4mm,
    every node/.append style={execute at begin node={\hyphenpenalty=10000}},
    box/.style={draw, rounded corners=2pt, align=center, inner sep=2pt},
    note/.style={align=center, text=black!70, font=\tiny}]
  \node[box, text width=15mm] (r1) {Agent 1};
  \node[box, text width=15mm, below=1.2mm of r1] (r2) {Agent 2};
  \node[box, text width=15mm, below=4mm of r2] (r3) {Agent $k$};
  \foreach \t in {0.25, 0.5, 0.75}
    \fill ($(r2.south)!\t!(r3.north)$) circle (0.4pt);
  \coordinate (mid) at ($(r1.north)!0.5!(r3.south)$);
  \node[box, text width=11mm, minimum height=12mm, left=of r1.west |- mid] (items) {Items};
  \node[box, text width=20mm, minimum height=12mm, right=8mm of r1.east |- mid] (gate)
    {Rule $\pi$:\\pool $A=1$\\or not $A=0$};
  \node[box, text width=20mm, right=9mm of gate, yshift=6mm] (judge) {Costly assessor:\\judgment $Y$};
  \node[box, text width=20mm, right=9mm of gate, yshift=-6mm] (none) {Not judged};
  \foreach \r in {r1, r2, r3} {
    \draw[->] (items.east) -- (\r.west);
    \draw[->] (\r.east) -- (gate.west);
  }
  \draw[->] (gate.east) -- (judge.west);
  \draw[->] (gate.east) -- (none.west);
  \draw[->, dashed, gray] (none.north) -- node[right, note] {audit} (judge.south);
  \draw[dashed, gray, rounded corners=2pt] ([shift={(-1mm,1mm)}]r1.north west)
    rectangle ([shift={(1mm,-1mm)}]r3.south east);
  \node[note, below=1.5mm of r3] {agreement: consensus only\\(Thm.~\ref{thm:agreement})};
  \node[note, above=0.5mm of judge] {acceptance: precision (Prop.~\ref{prop:acceptance})\\
    overlap: bounds (Thm.~\ref{thm:disagreement}, Prop.~\ref{prop:next})};
  \node[note, below=0.5mm of none] {relevant but unjudged $H$:\\audits estimate it (Prop.~\ref{prop:audit})};
\end{tikzpicture}
  \caption{A multi-agent screening system. Grey notes: what each kind of evidence shows.}
  \label{fig:pipeline}
\end{figure}
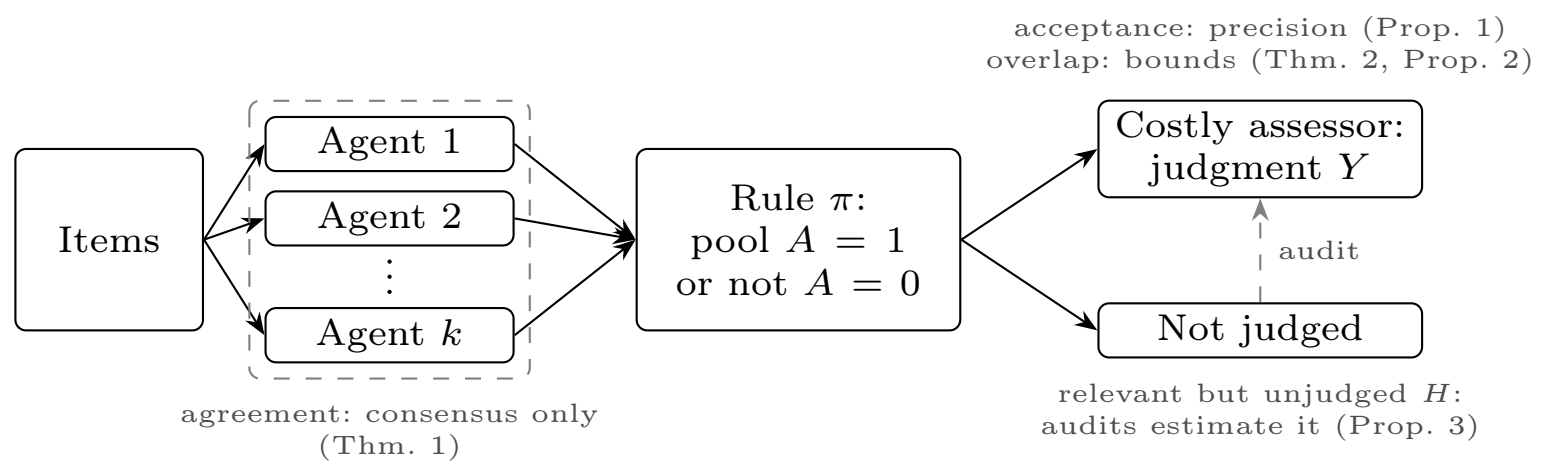

A \emph{multi-agent screening} system (Figure~\ref{fig:pipeline}) processes a fixed collection of items, such as documents, code changes or papers. 
Each of $k$ agents returns an output $O_i$, and $O=(O_1,\dots,O_k)$. 
A rule $A=\pi(O)$ \emph{retrieves} items with $A=1$ to the \emph{pool}, which is then submitted to the assessor. 
An item is \emph{relevant}, $Y=1$, if the assessor would accept it under a fixed standard (as in Cranfield-style evaluation), and $Y=0$ otherwise; we observe $Y$ only in the pool. 
Note that only the assessor accepts: agents merely pass items on. 
Probabilities refer to an item drawn uniformly from the collection: 
$a=\Prob(A=1)$ is the share of the collection in the pool, $p=\Prob(Y=1 \mid A=1)$ the acceptance rate in the pool (which is its precision), 
and $H=\Prob(Y=1,A=0)$ the proportion of the collection that is relevant but unjudged.
So recall is $r=ap/(ap+H)$, and $H$ is the unknown.

We consider four kinds of evidence: 
\emph{agreement}, the joint distribution of $O$;
\emph{acceptance}, the assessor's accept-or-reject judgments on the pool; 
\emph{overlap}, those
judgments split by who nominated each item; and 
\emph{audit}, judgments on a random sample from outside the pool (an elusion test in e-discovery). 
We give \emph{tight intervals}: the values that fit the data and our assumptions, and no others.

\section{What Each Kind of Evidence Can Tell Us}
\label{sec:limits}

\paragraph{Agreement: consensus, not correctness.}
Agents built on similar models err together~\cite{kim2025correlated}. 
Model this as a shared representation: a latent $Z$ with $Y$ and $O$ conditionally independent given $Z$, so outputs reflect relevance only through $Z$. 
For a rule $A=\pi(O)$, let $g(z)=\Prob(A=1\mid Z=z)$ and $\eta(z)=\Prob(Y=1\mid Z=z)$.

\begin{theorem}[Agreement]\label{thm:agreement}
For every $k$ and every rule $\pi$:
\emph{(a)} the error of $\pi$, $\Prob(A\ne Y)$, is at least $\E\min\{\eta(Z),1-\eta(Z)\}$;
\emph{(b)} the distribution of $O$ does not depend on $\eta$, and as $\eta$ varies the accuracy of $\pi$ takes every value in $[\E\min\{g(Z),1-g(Z)\},\,\E\max\{g(Z),1-g(Z)\}]$, which is $[0,1]$ if $A$ is a function of $Z$.
\end{theorem}
\begin{proof}
Fix $z$ and write $g,\eta$ for $g(z),\eta(z)$. 
Given $Z=z$, $A$ and $Y$ are independent, so $A=Y$ with probability $g\eta+(1-g)(1-\eta)\le\max\{\eta,1-\eta\}$, and $A\ne Y$ with probability at least $\min\{\eta,1-\eta\}$; average over $Z$.
The distribution of $(Z,O)$, and so $g$, is the same for every $\eta(\cdot)$, and the accuracy given $Z=z$ runs linearly from $1-g$ at $\eta=0$ to $g$ at $\eta=1$: pick $\eta(z)\in\{0,1\}$ for each end, and mix.
\qed
\end{proof}

For a representation that stays fixed as agents are added, 
(a) is an error floor at every $k$, positive unless $Z$ determines $Y$, and 
(b) says no statistic of the outputs can tell a right consensus from a wrong one. 
Label-free methods such as Dawid--Skene identify error rates against a discrete $Z$, up to relabelling, under conditional independence and rank conditions~\cite{dawid1979maximum,allman2009identifiability}; 
reading $Z$ as $Y$ is an assumption outputs cannot test, and shared-confounder models need assumptions or heuristics to say which factor is quality~\cite{zhao2026care}. 
With additive scores and centred errors, comparisons reveal agents' relative biases but not their shared bias~\cite{barooah2008estimation,balzano2007blind}.

\paragraph{Acceptance: precision, not recall.}
Acceptance estimates $p$, not $H$: the \emph{selective-labels} problem~\cite{lakkaraju2017selective}. 
Econometrics calls such bounds \emph{partial identification}~\cite{manski2003partial}.

\begin{proposition}[Acceptance]\label{prop:acceptance}
For any $\pi$ and $k$ with $ap>0$, the tight interval for recall is $[ap/(ap+1-a),\,1]$, since $r=ap/(ap+H)$ decreases in $H$ and the data allow any $H\in[0,1-a]$.
\end{proposition}

\paragraph{Overlap: an upper bound on recall.}
Let each agent nominate, $A_i\in\{0,1\}$, and the assessor judge the union $A=\max_iA_i$, as in pooled evaluation. 
Judgments show who found each relevant item. 
Take two disjoint non-empty groups of agents, let $B$ and $B'$ indicate a nomination by each and $C$ that all other agents stay silent, and write $h_{10}=\Prob(Y=1,C,B=1,B'=0)$, and likewise $h_{01}$ and $h_{11}$; all are observed in the pool. 
Fix a family of such splits and assume on each \emph{positive dependence}: $\mathrm{Cov}(B,B'\mid Y=1,C)\ge0$,
so that among relevant items all other agents miss, a nomination by one group makes one by the other no less likely, and $h_{11}>0$. 
By Chebyshev's association inequality the covariance condition holds on every split if agents are mutually independent given $(\Theta,Y=1)$ for a scalar item findability $\Theta$ and each $\Prob(A_i=1\mid\Theta,Y=1)$ is non-decreasing in $\Theta$.

\begin{theorem}[Overlap]\label{thm:disagreement}
Let $L$ be the largest $h_{10}h_{01}/h_{11}$ over the family. 
Then $H\ge L$. 
If $L\le1-a$, the tight interval for recall under these assumptions is $\bigl[ap/(ap+1-a),\ ap/(ap+L)\bigr]$; 
if $L>1-a$, the data refute them.
\end{theorem}
\begin{proof}
The joint shares of $(B,B')$ with $Y=1$ and $C$ form a $2\times2$ table whose unobserved cell, found by neither group, is $H$, and the conditional covariance has the sign of $Hh_{11}-h_{10}h_{01}$. 
Any $H\in[L,1-a]$ satisfies these inequalities and fits the data, since items with $A=0$ have no judgments.
\qed
\end{proof}

Little overlap means many misses. 
The ratio $h_{10}h_{01}/h_{11}$ is the Lincoln--Petersen \emph{capture-recapture} quantity~\cite{fienberg1972multiple}; without an assumption the unobserved cell cannot be determined, however many lists there are~\cite{aleshin2024central}. The theorem adapts the odds-ratio restrictions of Sun et al.~\cite{sun2023dependence} with lower bound one, adding the cap $H\le1-a$. 
Items invisible to every agent are consistent with positive dependence, so overlap cannot raise the lower end. 
Specialists covering different parts of the collection can violate the assumption; then apply the bound within a specialism where it holds. 
Sampled pools need weighting and add variance.

\paragraph{The next agent.}
Let $f_1=\Prob(Y=1$, exactly one of the first $k$ agents nominates$)$.

\begin{proposition}[Next agent]\label{prop:next}
If $A_1,\dots,A_{k+1}$ are exchangeable given $Y=1$ (relabelling agents changes nothing), agent $k+1$ adds new relevant items worth at most a share $f_1/k$ of the collection.
\end{proposition}
\begin{proof}
Among relevant items, ``agent $k+1$ nominates and $1,\dots,k$ do not'' implies ``agent $k+1$ nominates and $1,\dots,k-1$ do not'', which by exchangeability is as likely as ``agent $k$ alone among the first $k$ nominates'', a share $f_1/k$.
\qed
\end{proof}
Here exchangeability replaces the independent sampling behind incidence-based Good--Turing estimates~\cite{good1953population,chao2017thirty}. 
An agent of a new kind is not covered.

\paragraph{Audit: closing the gap.}
Only judgments outside the pool pin down $H$ without assumptions on the agents.

\begin{proposition}[Audit]\label{prop:audit}
Let $\hat H$ be $1-a$ times the relevant share of $n_0$ items drawn uniformly without replacement outside the pool. 
Then $|\hat H-H|\le\epsilon$ with probability $\ge1-\delta$
once $n_0\ge(1-a)^2\ln(2/\delta)/(2\epsilon^2)$, for any $k$ (Hoeffding~\cite{hoeffding1963probability}).
\end{proposition}
This is sufficient, not necessary: for 1000 items, $a=0.2$, $\delta=0.05$ and $\pm0.05$, it
asks for 473 audits, where an exact hypergeometric calculation needs 192.

\section{Worked Example: AI Reviewers and Human Experts}
\label{sec:peer-review}

Here items are paper submissions, agents are AI reviewers, and a paper is relevant if a human expert would accept it. 
At a hypothetical venue, three reviewers nominate 200 of 1000 submissions ($a=0.2$), and experts accept 120 ($p=0.6$). 
Nominations alone fit $p=0$ or $1$ as well as $0.6$ unless reviewers err independently given acceptability (Theorem~\ref{thm:agreement}).
Acceptance alone gives $r\in[13\%,100\%]$ (Proposition~\ref{prop:acceptance}). 
Of the 120, each reviewer alone nominated 30, each pair alone 6, and all three 12.
With reviewer~3 silent, reviewers 1 and 2 give counts 30, 30 and 6, so at least $30{\times}30/6=150$ acceptable papers were missed if reviewers are positively dependent, and $r\in[13\%,44\%]$ (Theorem~\ref{thm:disagreement}).
The venue can buy 200 more expert reviews. 
A 4th reviewer, if interchangeable, adds at most $90/3=30$ acceptable papers (Proposition~\ref{prop:next}). 
An audit of 192 of the other 800 submissions estimates $H$ to $\pm0.05$ with 95\% confidence (Proposition~\ref{prop:audit}) and, under positive dependence, finds at least $192{\times}150/800=36$ such papers in expectation. 
It finds more and alone certifies the misses but uses nearly all the budget; the 4th reviewer needs reviews only for its new nominations, which could be far fewer.

\section{Take-Home Messages}
\label{sec:rules}

Rate agents on judgments, not agreement, which can reward shared bias. 
Report precision, and recall as an interval: a single number claims too much. 
Let agents nominate separately: judge their union and log who found each item, since a threshold on an aggregate score can leave nominations unjudged. 
Compare the next agent with an audit before buying either, and budget audits by the accuracy wanted, not the number of agents; an audit can also reject positive dependence.

\bibliographystyle{splncs04}
\bibliography{ecir/references}

\end{document}